# Large barocaloric effects at low pressures in natural rubber

E. O. Usuda[1], N. M. Bom[1] and A. M. G. Carvalho[1,a]

[1]*Laboratório Nacional de Luz Síncrotron (LNLS), Centro Nacional de Pesquisa em Energia e Materiais (CNPEM), CEP 13083-970, Campinas, São Paulo, Brazil.*

[a]*Corresponding author. E-mail: alexandre.carvalho@lnls.br. Orcid: 0000-0001-6439-414X*

## ABSTRACT

Barocaloric effect in vulcanized natural rubber (V-NR) has been investigated. Direct measurements of the temperature change ($\Delta T$) around room temperature (283-333 K) resulted in large values, above 10 K, for a pressure change of 173 MPa. A power law was proposed to fit $\Delta T$ as function of the maximum pressure, showing to be suitable for the barocaloric effect in V-NR. Strain was measured as a function of temperature at constant pressures in order to obtain the isothermal entropy change ($\Delta S_T$). At 293 K, we obtained a $\Delta S_T$ of 21 J.kg$^{-1}$K$^{-1}$ for a pressure change of only 43.4 MPa. The results presented in this work are compared with those reported recently for PVDF-TrFE-CTFE polymer, showing a better barocaloric performance for V-NR in similar temperature and pressure ranges. These findings evidence the high potential of V-NR for application in solid-state refrigeration based on confined compression, opening new possibilities for i-caloric materials.

## 1. Introduction

Solid-state cooling based on *i*-caloric effects has experienced large research interest in the last decades, due to priorities in energy-saving and eco-friendliness [1–4]. The *i*-caloric effects (the "*i*" stands for intensive thermodynamic variables) are categorized as magnetocaloric effect (*h*-CE), electrocaloric effect (*e*-CE) and mechanocaloric effect (σ-CE), depending on the nature of the external field applied to the material (magnetic field, electrical field or mechanical stress). σ-CE can be induced by uniaxial stress ($\sigma_e$-CE) or isostatic pressure ($\sigma_b$-CE), for instance. Despite the impressive scientific and technological advances reported for both *h*-CE [5–8] and *e*-CE [9,10], serious disadvantages are still hindering their use in commercial refrigeration devices. Magnetocaloric materials and permanent magnets are mainly based on rare-earth compounds, whose production is very costly and ecologically detrimental [11,12]. Moreover, large magnets are necessary in order to apply the high fields (2-5 T) required to induce significant *h*-CE values, representing a practical problem for commercial devices. In turn, electrocaloric materials present risk of breakdown at high electric fields and are limited to thin film materials [9,13]. Finally, cooling devices based on *h*-CE and *e*-CE still exhibit poor coefficients of performance (COP) [2,9,14].

σ-CE was the first *i*-caloric effect discovered, being detected by John Gough in natural rubber (NR) in the beginning of the nineteenth century [15], and further studied by Joule [16]. In the 1940s, Dart and co-authors [17,18] have shown some interesting results from stretching experiments with NR and other elastomers, achieving elastocaloric temperature changes higher than 10 K. Despite this pioneering, σ-CE is the least investigated *i*-caloric effect so far. In the last decade, nevertheless, there was a resurgence of mechanocaloric research, as a consequence of the observation of giant $\sigma_e$-CE and $\sigma_b$-CE in intermetallics [19–23]. Besides, elastocaloric cooling present the largest potential among *i*-caloric alternatives according to US Department of Energy [24], and barocaloric materials exhibit caloric efficiencies comparable to the best magnetocaloric materials [3]. Following this trend, Tušek and co-authors [25] recently demonstrated an elastocaloric regenerative heat pump, with COP values up to 7 and 15.3 K of temperature span. However, large uniaxial stresses are often required (~ 0.6-0.8

GPa) in order to achieve such large temperature variations in intermetallics. This issue leads to the investigation of soft materials. In this context, NR appears as the most prominent mechanocaloric material due to its favorable characteristics: (i) high caloric performance; (ii) high resilience; (iii) low cost; and iv) low environmental impact. Some recent papers have reported encouraging results of $\sigma_e$-CE for NR [26–29]. The large elastocaloric temperature variations observed in this material are attributed to the reversible alignment of its polymer chains [26] or to the strain-induced crystallization (SIC) occasioned by the stretching [27].

Regarding $\sigma_b$-CE, there is a lack of data in the literature for many materials classes [3]. In the specific case of polymers, this effect is only systematically studied in PVDF-TrFE-CTFE [30], to the best of our knowledge. Nevertheless, it was demonstrated that $\sigma_b$-CE presents the most promising $i$-caloric efficiencies, justifying research efforts in this direction [3]. Taking the above discussion into account, the present study systematically investigates $\sigma_b$-CE in NR at several temperatures and pressures, in order to explore the mechanocaloric possibilities of this material. The experimental setup used here was previously developed by our research group and it is described in detail in Ref. [31]. The results show large variations in $i$-caloric parameters (i.e., temperature and entropy) by applying moderate pressures on NR. The findings presented here open interesting perspectives for future developments in mechanocaloric research field, especially for polymers.

## 2. Experimental

The vulcanized natural rubber (V-NR) sample was prepared from a liquid pre-vulcanized latex (purchased from Siquiplas) and cast into a cylinder with dimensions of 19.5 mm (length) and 12 mm (diameter). We used a cylindrical plaster mold, with diameter of 12 mm and length of 30 mm, to prepare the V-NR sample. A liquid latex feeder was necessary to prevent the formation of cavities due to the shrinkage of the latex while drying. The liquid latex was poured into the mold until it was completely filled (including the feeder) and then it was allowed to dry for 48 hours at room temperature (around 23º C). The FTIR spectrum of the sample is shown in Figure 1. After this process, the material was cut into the desired dimensions. The sample's density, 902(7) kg.m$^{-3}$, was determined using a pycnometer. Before being used in the experiments, the V-NR sample was submitted to several barocaloric experiments within a wide range of temperature (280 - 335 K) and pressure (up to 173 MPa), including the measurements reported in the ref. [31].

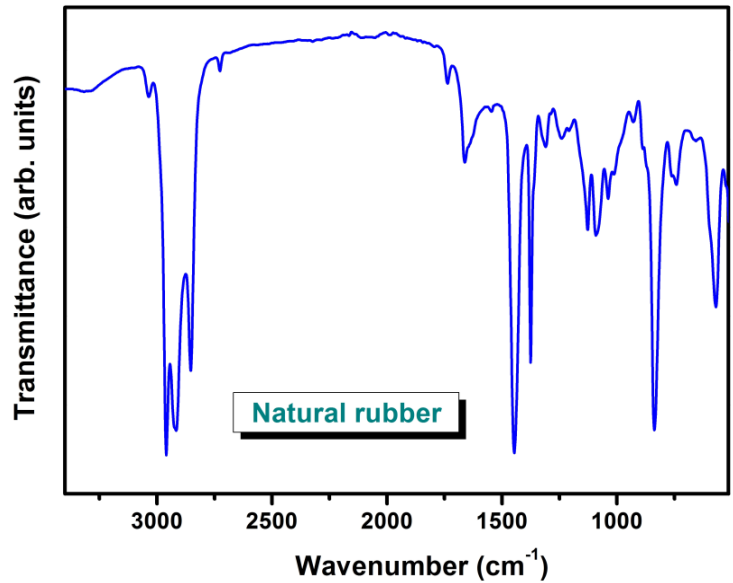

**Figure 1**. FTIR spectrum of the 12-mm-diameter sample of vulcanized natural rubber.

The experimental setup consists of a cylinder-shape carbon-steel chamber enveloped by a copper coil, in which a thermal fluid circulates pumped by a thermostatic bath (TE 184, Tecnal) for refrigerating/heating. A 12-mm-diameter piston moves through a cylindrical hole in the center of the chamber, where the sample is placed in. Below the sample, there is a bottom closure to hold the sample inside the chamber and to guide a thermocouple, whose tip is placed in the sample to measure its temperature in real time. A manual hydraulic press (P15500, Bonevau) is used to apply uniaxial load at the piston. Underneath the system, a load cell (3101C, ALFA Instrumentos) measures the contact force, which is used to indirectly calculate the pressure on the sample. Finally, a rod is attached to the piston to measure the displacement by a precise linear length gauge (METRO 2500, Heidenhain Co). For more information, a detailed view of the system is shown in Ref. [31].

$\sigma_b$-CE data were evaluated by two experimental methods: (1) measuring the temperature variation (ΔT) of the sample by applying pressure adiabatically (very short time interval); (2) measuring the strain variation of the sample while changing the temperature (ε *vs.* T), indirectly obtaining the isothermal entropy change ($\Delta S_T$). For the first method, different temperatures in the range of ~ 285-330 K were set in the thermostatic bath at a time, in such a way that the temperature values in the sample were the desired ones. Then, the sample was submitted to compressive stresses adiabatically, observing an immediately increasing in temperature of the sample. The final values of compressive stresses are in the range of 8.7-173 MPa. The load was kept constant until the temperature downs to the set point, and then the pressure was released adiabatically. This causes an abrupt decrease in the sample's temperature. Figure 2 shows the typical behavior of the temperature during these experiments for the applied pressures of 87(2) and 173(3) MPa at ~ 293 K, where the peak and the valley correspond to the adiabatic compression and decompression steps, respectively. This behavior is found in all *i*-caloric effects [32–34].

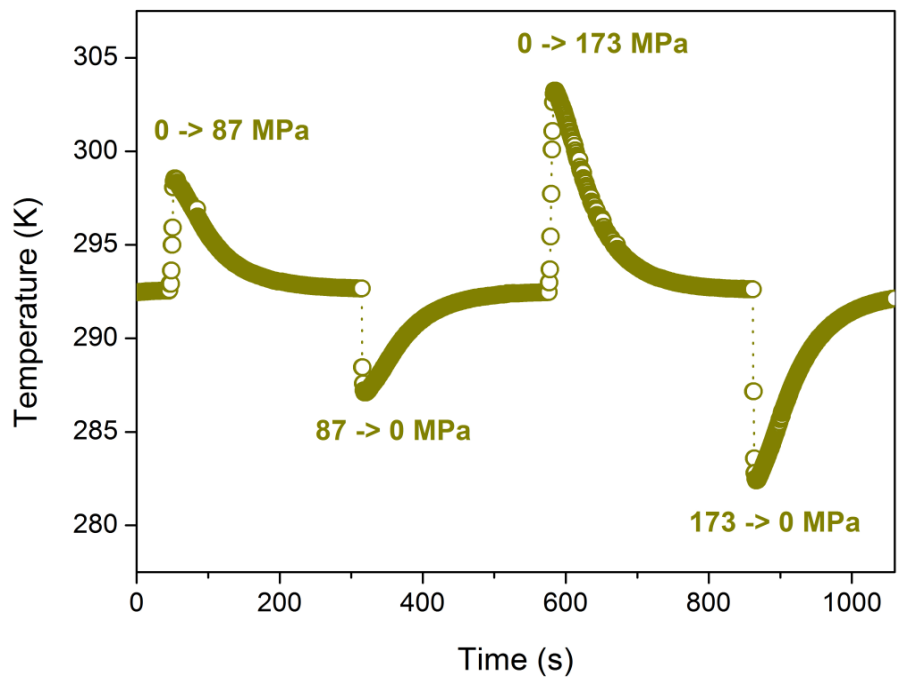

**Figure 2.** Typical temperature-time curve of V-NR for two pressures at ~ 293 K: 87(2) MPa and 173(3) MPa. The peaks represent the temperature changes when the pressures are applied; the valleys represent the temperature changes when the pressures are released.

## 3. Results and discussion

Figure 3 displays the modulus of the barocaloric temperature changes ($|\Delta T|$) obtained from the aforementioned experiments, as a function of the initial temperature at the following pressure variations: 26.0(5), 43.4(9), 87(2) and 173(3) MPa. Although $\Delta T$ data for compression processes are slightly larger, the results from decompression processes will be emphasized here, since releasing the pressure is a faster and more reproducible process than applying pressure in our apparatus, constituting a better approach to the adiabatic condition. It is possible to observe that all curves in decompression (Figure 3a) exhibit a similar qualitative trend: higher $|\Delta T|$ values are obtained as the initial temperature increases, and this effect is more pronounced for higher pressure changes. This behavior can possibly be assigned to the rearrangement of the V-NR polymer chains, analogously to which was observed in elastocaloric experiments. In the latter case, the degree of entropy decreases due to the alignment of polymer chains induced by stretching, leading to the increase of the temperature in V-NR [26]. For $\sigma_b$-CE, however, a detailed description of the physico-chemical mechanisms associated to the observed $\Delta T$ is still missing, and further studies are necessary to unravel this question. Moreover, no abrupt changes in $|\Delta T|$ are verified as temperature increases, indicating the absence of SIC or other phase transitions, differently from what was previously reported in mechanocaloric studies with V-NR [27,28] and PVDF-polymers [30]. Furthermore, around room temperature, the $\Delta T$ values observed in our data set for 173 MPa are significantly higher than those obtained for $\sigma_b$-CE in PVDF-TrFE-CTFE at similar pressures, as showed in Table 1. Regarding the normalized temperature change ($\Delta T/\Delta\sigma$), the difference is even higher, reaching 60.7 K/GPa in V-NR at 303 K, almost twice the value exhibited by the PVDF-TrFE-CTFE ($\approx$ 32.0 K/GPa). These results indicate the great barocaloric potential of the V-NR, suggesting an alternative way of exploring this material in mechanocaloric applications.

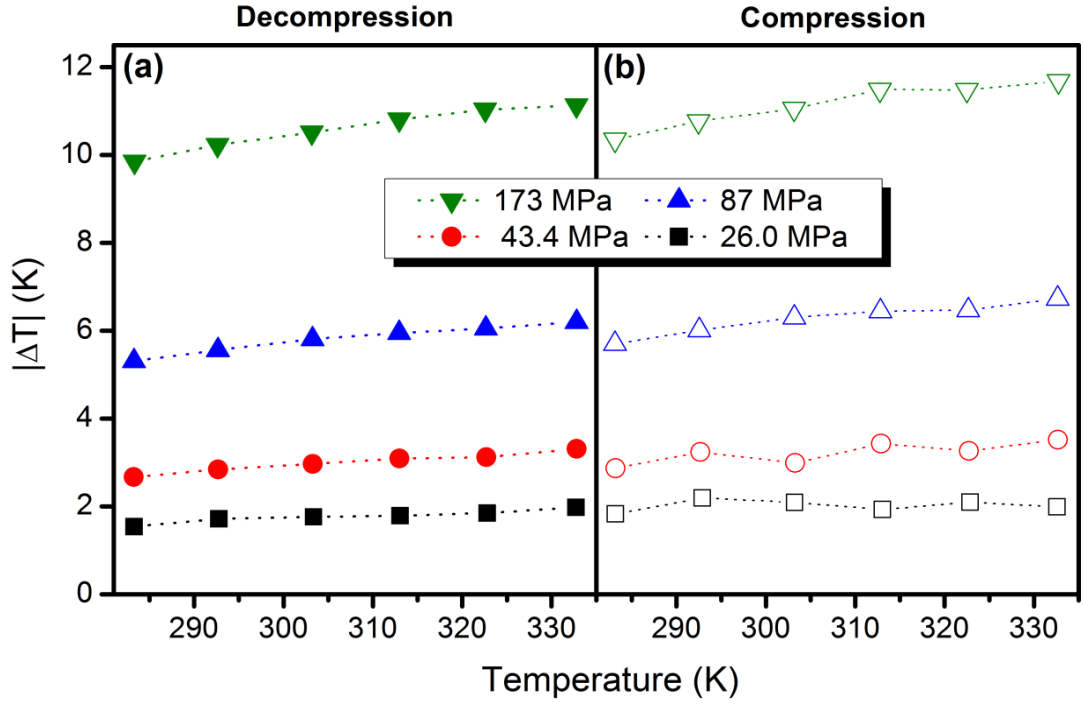

**Figure 3.** Temperature change (modulus) of V-NR as a function of initial temperature at the following pressure variations: 26.0(5) MPa (squares), 43.4(9) MPa (circles), 87(2) MPa (up triangles) and 173(3) MPa (down triangles).

**Table 1.** Barocaloric properties of V-NR and PVDF-TrFE-CTFE around room temperature. ΔT is the temperature change; ΔT/Δσ is the normalized temperature change; $\Delta S_T$ is the isothermal entropy change; $\Delta S_T \cdot \Delta T_S / \Delta \sigma$ is the normalized barocaloric power.

| Materials | T (K) | ΔT (K) | ΔT/Δσ (K.GPa$^{-1}$) | $\Delta S_T$ (J.kg$^{-1}$.K$^{-1}$) | $\Delta S_T \cdot \Delta T_S / \Delta \sigma$ (J.kg$^{-1}$.GPa$^{-1}$) | References |
|---|---|---|---|---|---|---|
| **V-NR** | 293 | -10.24* | 59.2(1) | -21(4) | 1.4(3)×10$^3$ | |
| | 303 | -10.50* | 60.7(1) | -17(4) | 1.2(3)×10$^3$ | This work |
| | 313 | -10.81* | 62.5(1) | -14(3) | 1.0(3)×10$^3$ | |
| | | (Δσ = -173 MPa) | | (Δσ = 43.4 MPa) | (Δσ = 43.4 MPa) | |
| **PVDF-TrFE-CTFE** | 303 | ≈ -6.4** | ≈ 32.0 | ≈ -13.6 | ≈ 0.7×10$^3$ | [30] |
| | 313 | ≈ -6.7** | ≈ 33.5 | ≈ -15.2 | ≈1.0×10$^3$ | |
| | | (Δσ = -200 MPa) | | (Δσ = 50 MPa) | (Δσ = 50 MPa) | |

*\* Direct measurement*
*\*\* Indirect method*

Figure 4 shows three curves for |ΔT| (at 283, 303 and 333 K) as function of the maximum released pressure ($\sigma_{max}$). According to previous works in *i*-caloric effects [35–37] the temperature change follows a specific proportionality with the intensive thermodynamic variables. Oesterreicher and Parker [35] and Franco *et al.* [36] show a power law for *h*-CE, where ΔT is proportional to the magnetic field (ΔT ∝ $H^{2/3}$). A power law is also reported by Lu *et al.* [37] for *e*-CE in ferroelectric polymers, where ΔT is proportional to low electric fields (ΔT ∝ $E^2$). Considering our results, we assume that, in the case of $\sigma_b$-CE for natural rubber, |ΔT| and pressure also follow a power law:

$$|\Delta T(T, \sigma_{max})| = a\sigma_{max}^n \qquad (1)$$

where *a* is the constant of proportionality, $\sigma_{max}$ is the maximum value of the released pressure and *n* is an exponent. The following values were obtained by fitting the data (Figure 4) using equation 1: for T = 283 K, *a* = 51(3) K.GPa$^{-n}$ and *n* = 0.94(3); for T = 303 K, *a* = 52(3) K.GPa$^{-n}$ and *n* = 0.91(3); for T = 333 K, *a* = 52(2) K.GPa$^{-n}$ and *n* = 0.88(2). The fitting curves are shown in Figure 4 by the solid lines. There is a good agreement between the experimental data and the simulated curves, showing that the power law (equation 1) is suitable for $\sigma_b$-CE as well. Besides, it was observed that the coefficients present slight variations as function of the initial temperature. Increasing the temperature, the constant of proportionality *a* increases and the exponent *n* decreases.

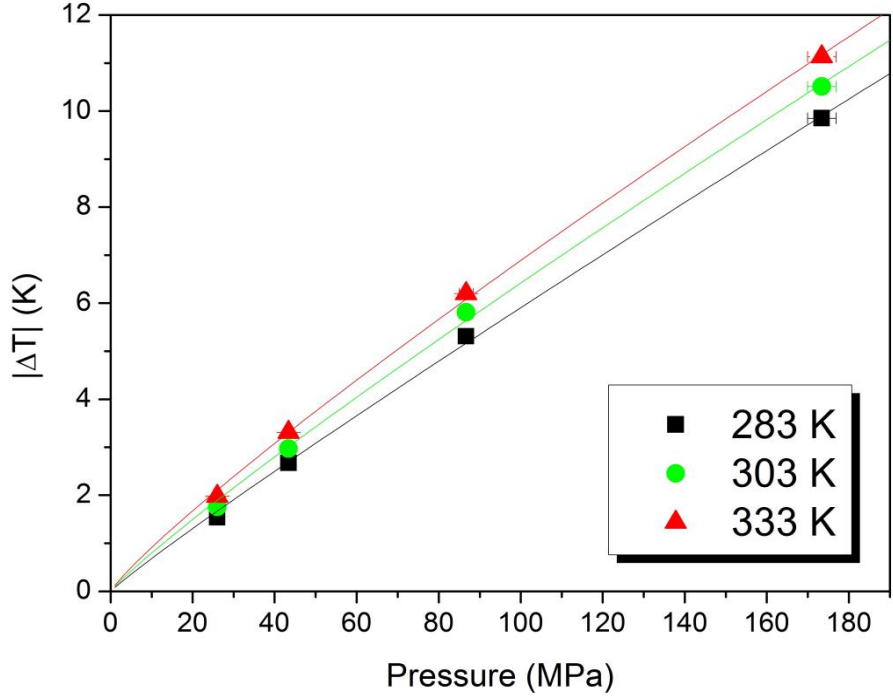

**Figure 4.** Temperature change (modulus) as a function of the released pressure for three initial temperatures: 283 K (squares), 303 K (circles) and 333 K (triangles).

ε vs. T curves (Figure 5) were measured for the determination of the $\Delta S_T$ in V-NR. The temperature of the sample was firstly set around 330 K for all applied pressures. Compressive stresses in the range of 0-43 MPa were applied at a time and kept constant (isobaric processes). The temperature was varied between ~ 330 K and ~ 285 K continuously, performing the following temperature cycle: 330 K -> 285 K -> 330 K -> 285 K. Figure 5 displays the final part of the isobaric cycle (285 K -> 330 K -> 285 K) for 8.7(2) and 26.0(6) MPa. Systematic errors (an offset) of 0.002 and random errors of 0.001 are estimated. The measured ε vs. T curves allow to indirectly quantify $\Delta S_T$ by using the following thermodynamic relation [30,38,39]:

$$\Delta S_T(T, \Delta\sigma) = -\frac{1}{\rho_0} \int_{\sigma_1}^{\sigma_2} \left(\frac{\partial \varepsilon}{\partial T}\right)_\sigma d\sigma \qquad (2)$$

where σ and $\rho_0$ are the compressive stress and the density of the sample at atmospheric pressure, respectively.

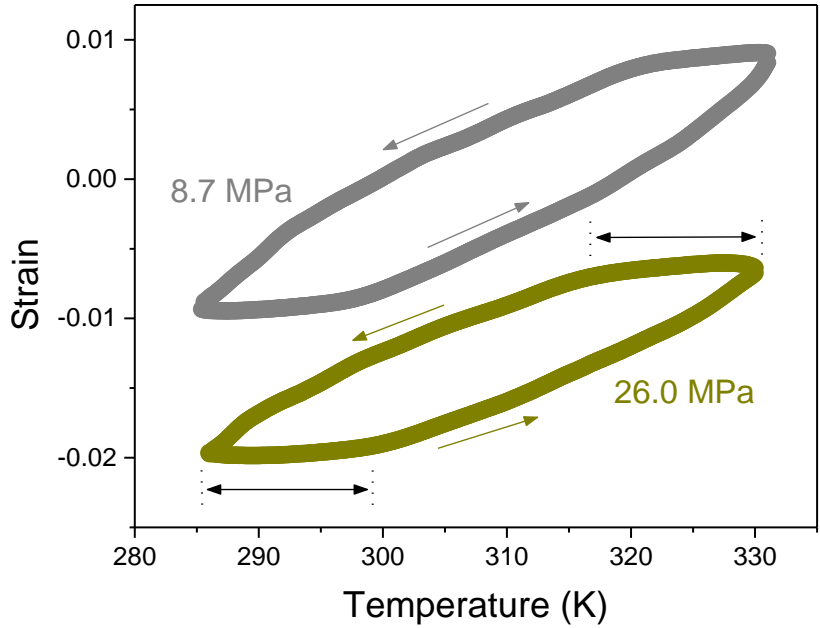

**Figure 5.** Strain as a function of temperature for V-NR at constant pressures of 8.7(2) and 26.0(6) MPa. The curves were smoothed using Origin® software.

It is clear that, in the beginning of the heating and cooling processes, the derivative dε/dT is smaller than the rest of the cycle. In Figure 5, the dotted lines mark these ranges of lower derivative in the curve for σ = 26.0 MPa. These ranges are observed also for σ = 8.7 MPa and for all isobaric curves measured up to 43.4 MPa (not shown). Since $\Delta S_T$ depends on the derivative dε/dT, these ranges of lower derivative should result in lower $\Delta S_T$ values, which is actually observed.

The $\Delta S_T$ values obtained from equation 2 are plotted as a function of temperature on heating and on cooling (Figure 6). The estimated errors for $\Delta S_T$ on heating is 1 J.kg$^{-1}$.K$^{-1}$ + 5%. For $\Delta S_T$ data on cooling, the estimated error is higher (1 J.kg$^{-1}$.K$^{-1}$ + 15%), so the error bars are omitted for sake of clarity. At 293 K, we observe the large $\Delta S_T$ of 21(4) J.kg$^{-1}$.K$^{-1}$ for the cooling curve, with no observable maximum. Considering $\Delta S_T$ values on cooling obtained for the pressure change of 43.4 MPa for V-NR and comparing with $\Delta S_T$ values for the pressure change of 50 MPa for PVDF-TrFE-CTFE (Table 1), we see that, around room temperature, the entropic $\sigma_b$-CE values are similar (taking the estimated errors into account). For T = 293 K and Δσ = 43.4 MPa, the COP value was estimated at 11(2) for V-NR.

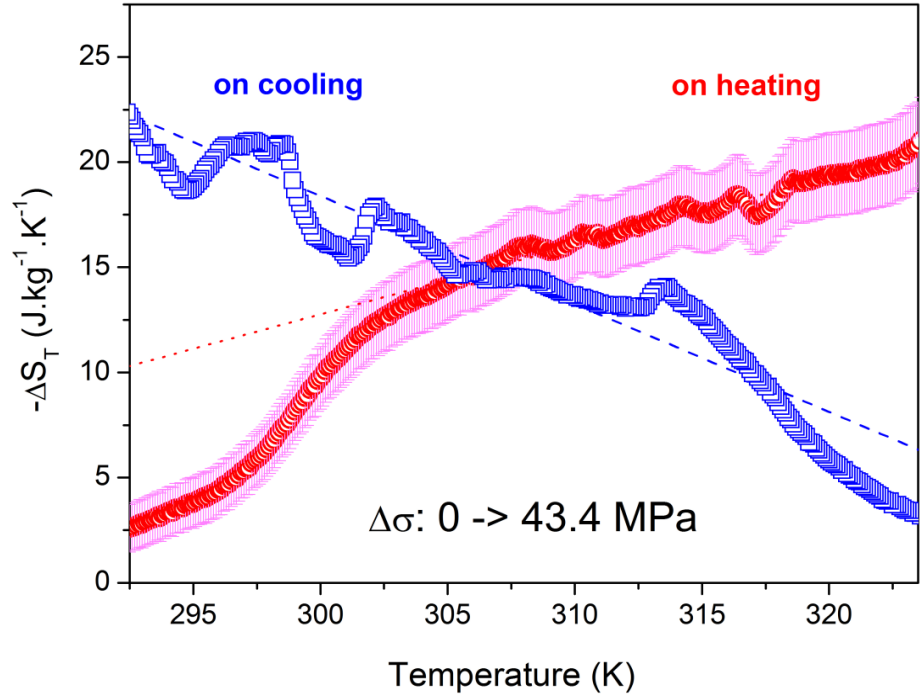

**Figure 6.** Entropy variation as function of temperature in V-NR within the 0-43.4 MPa pressure range for heating and cooling processes. The error bars for the cooling curve are not shown for better visualization of the results.

The dotted and dashed lines in Figure 6 represent linear fits, excluding the temperature ranges of lower dε/dT in ε vs. T curves. We suppose that the behavior of the $\Delta S_T$ curve on heating, for instance, should be closer to the dotted line if we reduce enough the lowest temperature of the cycle (down to ~ 240 K, for example). Since ε vs. T curves present hysteretic behavior, we assume those processes (heating and cooling) present irreversibility. Therefore, it is expected that the isothermal entropy change presents a component of irreversibility and $\Delta S_T$ values should depend on the cycle (the temperature range, for instance). The effects of the irreversibility in $\Delta S_T$ for *h*-CE have already been studied experimentally since last decade [40] and more recently for $\sigma_b$-CE [41].

In order to compare different barocaloric materials, we propose here the use of the *normalized barocaloric power* ($N\sigma_b CP$). This quantity depends on the temperature and gives us an estimation of the barocaloric potential at a certain temperature. We define the normalized barocaloric power as:

$$N\sigma_b CP(T) \equiv \frac{\Delta S_T(T) \cdot \Delta T_S(T)}{\Delta \sigma} \quad (3)$$

where $\Delta T_S$ is the adiabatic temperature change. In the current paper, the measured $\Delta T$ values for V-NR are certainly very close to the real $\Delta T_S$ values, thus we can use them in equation 3. For three selected temperatures around room temperature, 313, 303 and 293 K (Table 1), we have $N\sigma_b CP$ of $1.0(3)\times 10^3$, $1.2(3)\times 10^3$ and $1.4(3)\times 10^3$ J.kg$^{-1}$.GPa$^{-1}$, respectively, for $\Delta\sigma$ = 43.4 MPa. Therefore, reducing the temperature (around room temperature), the normalized barocaloric power of V-NR tends to increase.

## 4. Conclusion

In summary, barocaloric effect in natural rubber was measured around room temperature at pressures up to 173 MPa, using a customized experimental apparatus previously developed by our group. The results show that the temperature change ($\Delta T$) increases as function of the initial temperature, especially for higher pressures, reaching large values for 173 MPa (up to 11.13 K at 333 K). This behavior is probably related to some rearrangement of polymer chains in natural rubber. The absence of abrupt changes in the curves suggests that no phase transitions are being induced in the material as consequence of the compression/decompression processes. Moreover, the fitting of $\Delta T$ vs. pressure curves shows that the measured data follow a power law model, in a similar way to what was observed for magnetocaloric and electrocaloric effects. The isothermal entropy change ($\Delta S_T$) was indirectly determined from the experimental ε vs. T curves, reaching the value of 21 J.kg$^{-1}$K$^{-1}$ at 293 K for $\Delta\sigma$ of 43.4 MPa. This value was obtained from the ε vs. T curve measured on cooling and $\Delta S_T$ shows a tendency of increasing for lower temperatures. Finally, it was shown that all barocaloric quantities evaluated in the present study are similar or higher than those recently reported for PVDF-TrFE-CTFE. Based on the processes used to measure the temperature change (adiabatic and isobaric processes) and the barocaloric properties exhibited by V-NR, we envisage the application of the barocaloric effect in regenerative heat pumps using the Brayton cycle. Considering the practical advantages of natural rubber regarding low cost and low environmental impact, the present findings open encouraging perspectives for this material in the next generations of cooling devices.

**Acknowledgments**

The authors acknowledge financial support from FAPESP (project number 2012/03480-0), CNPq and CAPES. The authors also thank LNLS and CNPEM.

**References**

[1]     L. Mañosa, A. Planes, M. Acet, Advanced materials for solid-state refrigeration, J. Mater. Chem. A. 1 (2013) 4925–4936. doi:10.1039/b000000x.

[2]     X. Moya, S. Kar-Narayan, N.D. Mathur, Caloric materials near ferroic phase transitions, Nat. Mater. 13 (2014) 439–450. doi:10.1038/nmat3951.

[3]     S. Crossley, N.D. Mathur, X. Moya, New developments in caloric materials for cooling applications, AIP Adv. 5 (2015) 67153. doi:10.1063/1.4922871.

[4]     B. Lu, J. Liu, Mechanocaloric materials for solid-state cooling, Sci. Bull. 60 (2015) 1638. doi:10.1007/s11434-015-0898-5.

[5]     E. Brück, Developments in magnetocaloric refrigeration, J. Phys. D. Appl. Phys. 38 (2005) R381–R391. doi:10.1088/0022-3727/38/23/R01.

[6]     B. Yu, M. Liu, P.W. Egolf, A. Kitanovski, A review of magnetic refrigerator and heat pump prototypes built before the year 2010, Int. J. Refrig. 33 (2010) 1029–1060. doi:10.1016/j.ijrefrig.2010.04.002.

[7]     K.A. Gschneidner, V.K. Pecharsky, Thirty years of near room temperature magnetic cooling: Where we are today and future prospects, Int. J. Refrig. 31 (2008) 945–961. doi:10.1016/j.ijrefrig.2008.01.004.

[8]     V. Franco, J. Blázquez, B. Ingale, A. Conde, The Magnetocaloric Effect and Magnetic Refrigeration Near Room Temperature: Materials and Models, Annu. Rev. Mater. Res. 42 (2012) 305–342. doi:10.1146/annurev-matsci-062910-100356.

[9]     M. Valant, Electrocaloric materials for future solid-state refrigeration technologies, Prog. Mater. Sci. 57 (2012) 980–1009. doi:10.1016/j.pmatsci.2012.02.001.

[10]    J.F. Scott, Electrocaloric Materials, Annu. Rev. Mater. Res. 41 (2011) 229–240.

doi:10.1146/annurev-matsci-062910-100341.

[11] Z. Xie, G. Sebald, D. Guyomar, Comparison of caloric effects in view of environment and application, (2016) arXiv:1604.04479. https://arxiv.org/abs/1604.04479.

[12] J. Navarro, F. Zhao, Life-cycle assessment of the production of rare-earth elements for energy applications: A review, Front. Energy Res. 2 (2014) 1–17. doi:10.3389/fenrg.2014.00045.

[13] X. Yin, J.F. Capsal, D. Guyomar, A comprehensive investigation of poly(vinylidene fluoride-trifluoroethylene-chlorofluoroethylene) terpolymer nanocomposites with carbon black for electrostrictive applications, Appl. Phys. Lett. 104 (2014) 52913. doi:10.1063/1.4864160.

[14] M. Ozbolt, A. Kitanovski, J. Tusek, A. Poredos, Electrocaloric refrigeration: Thermodynamics, state of the art and future perspectives, Int. J. Refrig. 40 (2014) 174–188. doi:10.1016/j.ijrefrig.2013.11.007.

[15] J.A. Gough, A Description of a property of Caoutchouc or Indian Rubber, Mem. Lit. Phyiosophical Soc. Manchester. 1 (1805) 288–295. http://www.biodiversitylibrary.org/bibliography/49075.

[16] J.P. Joule, On some thermodynamic properties of solids, Phil. Trans. R. Soc. Lond. 149 (1859) 91–131.

[17] S.L. Dart, R.L. Anthony, E. Guth, Rise of temperature on fast stretching of synthetic and natural rubbers, Ind. Eng. Chem. 34 (1942) 1340–1342. doi:10.5254/1.3546779.

[18] S.L. Dart, E. Guth, Rise of temperature on fast stretching of butyl rubber, Rubber Chem. Technol. 18 (1945) 803–816. doi:10.5254/1.3546779.

[19] E. Bonnot, R. Romero, L. Mañosa, E. Vives, A. Planes, Elastocaloric effect associated with the martensitic transition in shape-memory alloys, Phys. Rev. Lett. 100 (2008) 125901. doi:10.1103/PhysRevLett.100.125901.

[20] J. Cui, Y. Wu, J. Muehlbauer, Y. Hwang, R. Radermacher, S. Fackler, M. Wuttig, I. Takeuchi, Demonstration of high efficiency elastocaloric cooling with large delta T using NiTi wires, Appl. Phys. Lett. 101 (2012) 73904. doi:10.1063/1.4746257.

[21] J. Tušek, K. Engelbrecht, L.P. Mikkelsen, N. Pryds, Elastocaloric effect of Ni-Ti wire for application in a cooling device, J. Appl. Phys. 117 (2015) 124901. doi:10.1063/1.4913878.

[22] L. Mañosa, D. González-alonso, A. Planes, E. Bonnot, M. Barrio, J. Tamarit, S. Aksoy, M. Acet, Giant solid-state barocaloric effect in the Ni-Mn-In magnetic shape-memory alloy, Nat. Mater. 9 (2010) 478–481. doi:10.1038/nmat2731.

[23] E. Stern-Taulats, A. Planes, P. Lloveras, M. Barrio, J.L. Tamarit, S. Pramanick, S. Majumdar, C. Frontera, L. Mañosa, Barocaloric and magnetocaloric effects in $Fe_{49}Rh_{51}$, Phys. Rev. B - Condens. Matter Mater. Phys. 89 (2014) 214105. doi:10.1103/PhysRevB.89.214105.

[24] W. Goetzler, R. Zogg, J. Young, C. Johnson, Energy savings potential and RD&D oppportunities for non- vapor-compression HVAC technologies, 2014.

[25] J. Tušek, K. Engelbrecht, D. Eriksen, S. Dall'Olio, J. Tušek, N. Pryds, A regenerative elastocaloric heat pump, Nat. Energy. 1 (2016) 16134. doi:10.1038/nenergy.2016.134.

[26] D. Guyomar, Y. Li, G. Sebald, P. Cottinet, B. Ducharne, J. Capsal, Elastocaloric modeling of natural rubber, Appl. Therm. Eng. 57 (2013) 33–38. doi:10.1016/j.applthermaleng.2013.03.032.

[27] Z. Xie, G. Sebald, D. Guyomar, Z. Xie, G. Sebald, D. Guyomar, Elastocaloric effect dependence on pre-elongation in natural rubber Elastocaloric, Appl. Phys. Lett. 107 (2015) 81905. doi:10.1063/1.4929395.

[28] Z. Xie, G. Sebald, D. Guyomar, Comparison of direct and indirect measurement of the

elastocaloric effect in natural rubber, Appl. Phys. Lett. 108 (2016) 41901. doi:10.1063/1.4940378.

[29] G. Sebald, Z. Xie, D. Guyomar, Fatigue effect of elastocaloric properties in natural rubber, Philos. Trans. R. Soc. London A Math. Phys. Eng. Sci. 374 (2016) 439–450. doi:10.1098/rsta.2015.0302.

[30] S. Patel, A. Chauhan, R. Vaish, P. Thomas, Elastocaloric and barocaloric effects in polyvinylidene di-fluoride-based polymers, Appl. Phys. Lett. 108 (2016) 72903. doi:10.1063/1.4942000.

[31] N.M. Bom, E.O. Usuda, G.M. Guimarães, A.A. Coelho, A.M.G. Carvalho, Experimental setup for measuring the barocaloric effect in polymers: Application to natural rubber, Rev. Sci. Instrum. 88 (2017) 46103. http://arxiv.org/abs/1612.08638.

[32] A.M.G. Carvalho, C. Salazar Mejía, C.A. Ponte, L.E.L. Silva, J. Kaštil, J. Kamarád, A.M. Gomes, Adiabatic temperature change from non-adiabatic measurements, Appl. Phys. A. 122 (2016) 246. doi:10.1007/s00339-016-9775-6.

[33] Y. Jia, Y.S. Ju, Y. Jia, Y.S. Ju, A solid-state refrigerator based on the electrocaloric effect, Appl. Phys. Express. 100 (2012) 242901. doi:10.1063/1.4729038.

[34] A. Sakata, N. Suzuki, Y. Higashiura, T. Matsuo, T. Sato, Measurement of the mechanocaloric effect in rubber, J. Therm. 113 (2013) 1555–1563. doi:10.1007/s10973-013-3066-7.

[35] H. Oesterreicher, F.T. Parker, Magnetic cooling near Curie temperatures above 300 K, J. Appl. Phys. 55 (1984) 4334–4338. doi:10.1063/1.333046.

[36] V. Franco, A. Conde, J.M. Romero-Enrique, Y.I. Spichkin, V.I. Zverev, A.M. Tishin, Field dependence of the adiabatic temperature change in second order phase transition materials: Application to Gd, J. Appl. Phys. 106 (2009). doi:10.1063/1.3261843.

[37] S.G. Lu, B. Rozic, Q.M. Zhang, Z. Kutnjak, R. Pirc, Electrocaloric effect in ferroelectric polymers, Appl. Phys. A Mater. Sci. Process. 107 (2012) 559–566. doi:10.1007/s00339-012-6830-9.

[38] Y. Liu, I.C. Infante, X. Lou, L. Bellaiche, J.F. Scott, B. Dkhil, Giant room-temperature elastocaloric effect in ferroelectric ultrathin films, Adv. Mater. 26 (2014) 6132–6137. doi:10.1002/adma.201401935.

[39] M.M. Vopson, Theory of giant-caloric effects in multiferroic materials, J. Phys. D. Appl. Phys. 46 (2013) 345304. doi:10.1088/0022-3727/46/34/345304.

[40] A.M.G. Carvalho, A.A. Coelho, S. Gama, P.J. von Ranke, C.S. Alves, Isothermal variation of the entropy (delta S) for the compound $Gd_5Ge_4$ under hydrostatic pressure, J. Appl. Phys. 104 (2008) 63915. doi:10.1063/1.2980040.

[41] S. Yuce, M. Barrio, B. Emre, E. Stern-Taulats, A. Planes, J.L. Tamarit, Y. Mudryk, K.A. Gschneidner, V.K. Pecharsky, L. Mañosa, Barocaloric effect in the magnetocaloric prototype $Gd_5Si_2Ge_2$, Appl. Phys. Lett. 101 (2012). doi:10.1063/1.4745920.